\documentclass[twocolumn]{aastex631}
\usepackage{comment} 
\usepackage{amsmath}
\usepackage{newtxtext,newtxmath}
\usepackage[T1]{fontenc}
\usepackage{graphicx}	% Including figure files
\usepackage{amsmath}	% Advanced maths commands
\usepackage{amssymb}	% Extra maths symbols

\usepackage{color,soul}
\usepackage{ulem}
\usepackage{amssymb}

\newcommand{\hess}{H.E.S.S.}

\submitjournal{ApJL}

\shorttitle{Pulsars as PeVatrons}
\shortauthors{de O\~na Wilhelmi et al.}
\graphicspath{{./}{figures/}}
\begin{document}

\title{On the potential of bright, young pulsars to power ultra-high gamma-ray sources}

\author[0000-0002-5401-0744]{Emma de O\~na Wilhelmi}
\affiliation{Deutsches Elektronen Synchrotron DESY, 15738 Zeuthen, Germany}
\email{emma.de.ona.wilhelmi@desy.de}
\author[0000-0002-3882-9477]{Rub\'en L\'opez-Coto}
\email{rlopezcoto@gmail.com}
\affiliation{Istituto Nazionale di Fisica Nucleare, Sezione di Padova, I-35131, Padova, Italy}
\affiliation{Instituto de Astrof\'isica de Andaluc\'ia, CSIC, 18080 Granada, Spain.}
\author[0000-0002-9881-8112]{Elena Amato}
\affiliation{INAF, Osservatorio Astrofisico di Arcetri, Largo E. Fermi 5, I-50125 Firenze, Italy}
\affiliation{Dipartimento di Fisica e Astronomia, Universit\`a di Firenze, Via Sansone 1,50019 Sesto Fiorentino (FI), Italy}
\email{elena.amato@inaf.it}
\author[0000-0003-1157-3915]{Felix Aharonian}
\affiliation{Dublin Institute for Advanced Studies, 31 Fitzwilliam Place, Dublin, Ireland}
\affiliation{Max-Planck-Institut f\"ur Kernphysik, P.O. Box 103980, D 69029 Heidelberg, Germany}
\email{Felix.Aharonian@mpi-hd.mpg.de}

\begin{abstract}
The recent discovery of a new population of ultra-high-energy gamma-ray sources with spectra extending beyond $100 \, \rm TeV$ revealed the presence of Galactic PeVatrons - cosmic-ray factories accelerating particles to PeV energies. These sources, except for the one associated with the Crab Nebula,  are not yet identified. With an extension of 1 degree or more, most of them contain several potential counterparts, including Supernova Remnants, young stellar clusters and Pulsar Wind Nebulae (PWNe), which can perform as PeVatrons and thus power the surrounding diffuse ultra-high energy gamma-ray structures. In the case of PWNe, gamma rays are produced by electrons, accelerated at the pulsar wind termination shock, through the inverse Compton scattering of 2.7 K CMB radiation. The high conversion efficiency of pulsar rotational power to relativistic electrons, combined with the short cooling timescales, allow gamma-ray luminosities up to the level of $L_\gamma \sim 0.1 \dot{E}$. The pulsar spin-down luminosity, $\dot E$, also determines the absolute maximum energy of individual photons: $E_{\rm \gamma~\rm max}\approx 0.9 \dot E_{36}^{0.65}~~\rm{PeV}$. This fundamental constraint dominates  
%($\rm B\sim 100\mu$G)}\elena{(why this change here? Losses are irrelevant for $B< 100 \mu$G)}, then  with $\dot{E} \lesssim 10^{37}\ \rm erg/s$
over the condition set by synchrotron energy losses of electrons for young PWNe with typical magnetic field of $\approx$100~$\mu$G with $\dot{E} \lesssim 10^{37}\ \rm erg/s$. We discuss the implications of $E_{\rm \gamma~\rm max}$ by comparing it with the highest energy photons reported by LHAASO from a dozen of ultra-high-energy sources. Whenever a PWN origin of the emission is possible, we use the LHAASO measurements to set upper limits on the nebular magnetic field.

\end{abstract}



\section{Introduction}

%New >100 TeV sources
%In agreement with many TeV PWNe candidates
%Pulsars and K-N - CMB
%In here, we review the possibilities based in first principles

%One of the drivers of the astroparticle physics field is the understanding of the origin of Cosmic Rays (CRs). In particular, the question about the origin of Galactic CRs, whose spectral energy distribution (SED) extends up to PeV energies (1 PeV = 10$^{15}$eV), has triggered numerous theoretical and experimental works (see \citealt{2019IJMPD..2830022G,2014NuPhS.256....9B,2014IJMPD..2330013A, 2013A&ARv..21...70B} for recent reviews). 

%\rlc{General: We call the magnetic field at the termination shock $B_{TS}$ and $B$, we should unify the notation}

The recent discovery of the LHAASO collaboration \citep{2021Natur.594...33C}, reporting the detection of a dozen sources with particle spectra reaching PeV (1 PeV = 10$^{15}$eV) energies, represents a major step towards the identification of the nature of the sources known as PeV accelerators, or {\it PeVatrons}. These observations are complemented in the gamma-ray regime by those from instruments sensitive in the 100\,GeV to 100\,TeV energy range, in particular by water Cherenkov instruments like HAWC and Tibet AS$\gamma$ \citep{2020ApJ...905...76A,PhysRevLett.126.141101}, and Imaging Atmospheric Cherenkov Telescopes arrays (IACT) such as \hess{}, MAGIC, and VERITAS \citep{2018A&A...612A...1H,2015JHEAp...5...30A,2015ICRC...34..771P}. In general, IACTs provide a superior angular resolution, that can be used to localize the emission regions more accurately and identify the accelerator type. The majority of the sources reported by LHAASO in \citealt{2021Natur.594...33C} are described by an extended gamma-ray emission, as large as $\sim$1$^{\rm o}$. These sources have, in almost all cases, a sub-100\,TeV counterpart. Interestingly, in this sub-100 TeV regime, more than 30\% of the sources detected in the Galactic plane have been associated with Pulsar Wind Nebulae (PWNe) \citep{2008ICRC....3.1341W,2018A&A...612A...2H}. These associations are based on spatial correlation with energetic pulsars, and spectral-morphological features connecting the usually extended ($\gtrsim0.2^{\rm o}$) gamma-ray emission with the pulsars. %Aside from the fact that pulsars, or actually PWNe, are the only source class in which PeV particles have been detected (\citealt{amatoolmi21} and references therein), 
Pulsars (or PWNe) are the only identified source class in which PeV particles have been detected: the Crab Nebula, associated with the young, very energetic pulsar PSR\,B0531+21, shows a synchrotron (steady and flaring) spectrum in the GeV regime that corresponds to PeV electrons \citep{2010ApJ...708.1254A,2011Sci...331..739A}, and it has also been recently detected up to an energy of 1.1~PeV by the LHAASO experiment \citep{2021Sci...373..425L}. Nevertheless, pulsars appear to be close to the absolute theoretical limit in terms of acceleration rate~\citep{1995NuPhS..39..193A,2021Sci...373..425L}.

%Even when pulsar are in principle capable of particle acceleration to PeV energies, they appear to be close to the absolute theoretical limit in terms of acceleration rate~\citep{1995NuPhS..39..193A}. 

%Pulsars (or PWNe) are the only identified source class in which PeV particles have been detected: the Crab Nebula, associated with the young, very energetic pulsar PSR\,B0531+21, shows a synchrotron (steady and flaring) spectrum in the GeV regime that corresponds to PeV electrons \citep{2010ApJ...708.1254A,2011Sci...331..739A}, and it has also been recently detected up to an energy of 1.1~PeV by the LHAASO experiment \citep{2021Sci...373..425L}. \tbd{Nevertheless, pulsars appear to be close to the absolute theoretical limit in terms of acceleration rate~\citep{1995NuPhS..39..193A,2021Sci...373..425L}.}

%Moreover, these energies should quickly dissipate via inverse Compton (IC) mechanism off-scattering the 2.7 K CMB in relativistic Klein-Nishina regime. These arguments have been put forward to favor hadronic mechanisms as the origin of the multi-TeV emission. 

In the following, we investigate the capability of energetic pulsars to power {\it PeVatrons}.
\begin{deluxetable*}{ccccccccc}

%% Keep a portrait orientation

%% Over-ride the default font size
%% Use Default (12pt)

%% Use \tablewidth{?pt} to over-ride the default table width.
%% If you are unhappy with the default look at the end of the
%% *.log file to see what the default was set at before adjusting
%% this value.

%% This is the title of the table.
\tablecaption{The table lists the LHAASO ultra-high energy sources, together with the bright, young pulsars located within 1$^{\rm o}$ of the LHAASO source and their characteristics. The two right-most columns display the maximum energy quoted by \citealt{2021Natur.594...33C} and its corresponding energy in electron, using the formulation in \citealt{2014ApJ...783..100K}.}

%% This command over-rides LaTeX's natural table count
%% and replaces it with this number.  LaTeX will increment 
%% all other tables after this table based on this number
\tablenum{1}

%% The \tablehead gives provides the column headers.  It
%% is currently set up so that the column labels are on the
%% top line and the units surrounded by ()s are in the 
%% bottom line.  You may add more header information by writing
%% another line between these lines. For each column that requries
%% extra information be sure to include a \colhead{text} command
%% and remember to end any extra lines with \\ and include the 
%% correct number of &s.
\tablehead{\colhead{LHAASO Source} & \colhead{Pulsar} & \colhead{Separation} & \colhead{$\dot{\rm E}$ } & \colhead{Age} & \colhead{Distance} & \colhead{Flux$_{\rm 100 TeV}$} & \colhead{E$_{\rm \gamma\ LHAASO}$} & \colhead{E$_{\rm e\ LHAASO}$} \\ 
\colhead{} & \colhead{} & \colhead{[deg]} & \colhead{$\times10^{36}$[erg/s]} & \colhead{[kyr]} & \colhead{[kpc]} & \colhead{[c. u.]} & \colhead{[PeV]} & \colhead{[PeV]} } 

%% All data must appear between the \startdata and \enddata commands
\startdata
J1825-1326 & J1826-1256 & 0.51 & 3.6 & 14.4 & 1.55 & 3.57 & 0.42 & 1.10 \\
 & B1823-13 & 0.16 & 2.8 & 21.4 & 3.61 & 3.57 & 0.42 & 1.10 \\
J1839-0545 & J1837-0604 & 0.61 & 2.0 & 33.8 & 4.77 & 0.70 & 0.21 & 0.65 \\
 & J1838-0537 & 0.25 & 6.0 & 4.9 & -- & 0.70 & 0.21 & 0.65 \\
J1843-0338 & J1841-0345 & 0.37 & 0.3 & 55.9 & 3.78 & 0.73 & 0.26 & 0.76 \\
 & J1844-0346 & 0.41 & 4.2 & 11.6 & -- & 0.73 & 0.26 & 0.76 \\
J1849-0003 & J1849-0001 & 0.10 & 9.8 & 43.1 & -- & 0.74 & 0.35 & 0.96 \\
J1908+0621 & J1907+0602 & 0.32 & 2.8 & 19.5 & 2.37 & 1.36 & 0.44 & 1.14 \\
 & J1907+0631 & 0.33 & 0.5 & 11.3 & 3.40 & 1.36 & 0.44 & 1.14 \\
J1929+1745 & J1925+1720 & 0.94 & 0.9 & 115.0 & 5.06 & 0.38 & 0.71 & 1.65 \\
 & J1928+1746 & 0.07 & 1.6 & 82.6 & 4.34 & 0.38 & 0.71 & 1.65 \\
J1956+2845 & J1954+2836 & 0.44 & 1.0 & 69.4 & 1.96 & 0.41 & 0.42 & 1.10 \\
 & J1958+2846 & 0.54 & 0.3 & 21.7 & 1.95 & 0.41 & 0.42 & 1.10 \\
J2018+3651 & J2021+3651 & 0.42 & 3.4 & 17.2 & 1.80 & 0.50 & 0.27 & 0.78 \\
J2032+4102 & J2032+4127 & 0.41 & 0.1 & 201.0 & 1.33 & 0.54 & 1.42 & 2.82 \\
J2108+5157 & &&&&&&\\
J2226+6057 & J2229+6114 & 0.38 & 22.0 & 10.5 & 3.00 & 1.05 & 0.57 & 1.39 \\
\enddata

%% Include any \tablenotetext{key}{text}, \tablerefs{ref list},
%% or \tablecomments{text} between the \enddata and 
%% \end{deluxetable} commands

%% No \tablecomments indicated

%% No \tablerefs indicated

\end{deluxetable*}

\section{Pulsars as effective PeV accelerators}

%{\bf Maximum energy}

\begin{deluxetable}{ccccc}

%\tablewidth{0pt}
\tablewidth{0.8\columnwidth}
%% If you are unhappy with the default look at the end of the
%% *.log file to see what the default was set at before adjusting
%% this value.

%% This is the title of the table.
\tablecaption{The table lists the LHAASO ultra-high energy sources and putative associated pulsars, with the corresponding constraints on the maximum energy, efficiency and magnetic field.}

%% This command over-rides LaTeX's natural table count
%% and replaces it with this number.  LaTeX will increment 
%% all other tables after this table based on this number
\tablenum{2}

%% The \tablehead gives provides the column headers.  It
%% is currently set up so that the column labels are on the
%% top line and the units surrounded by ()s are in the 
%% bottom line.  You may add more header information by writing
%% another line between these lines. For each column that requries
%% extra information be sure to include a \colhead{text} command
%% and remember to end any extra lines with \\ and include the 
%% correct number of &s.
\tablehead{\colhead{LHAASO Source} & \colhead{Pulsar} & \colhead{E$_{\rm \gamma\ max}$} & \colhead{E$_{\rm max}$ }  & \colhead{B$_{\rm max}$} \\ 
\colhead{} & \colhead{} & \colhead{[PeV]} & \colhead{[PeV]} &  \colhead{[$\mu$G]} } 

%% All data must appear between the \startdata and \enddata commands
\startdata
J1825-1326 & J1826-1256 & 2.06 & 3.79 & 38\\
           & B1823-13 & 1.77 & 3.35 & 14\\
J1839-0545 & J1837-0604 & 1.44 & 2.83 & 33\\
           & J1838-0537 & 2.78 & 4.90 & $\gg$100\\
J1843-0338 & J1841-0345 & 0.41 & 1.04 & 12\\
           & J1844-0346 & 2.25 & 4.10 & $\gg$100\\
J1849-0003 & J1849-0001 & 3.71 & 6.26 & $\gg$100\\
J1908+0621 & J1907+0602 & 1.77 & 3.35 & 30\\
            & J1907+0631 & 0.63 & 1.46 & 9\\
J1929+1745 & J1925+1720 & 0.91 & 1.95 & 9\\
            & J1928+1746 & 1.26 & 2.53 & 14\\
J1956+2845 & J1954+2836 & 0.94 & 2.00 & 37\\
           & J1958+2846 & 0.47 & 1.17 & 22\\
J2018+3651 & J2021+3651 & 1.99 & 3.69 & 102\\
J2032+4102 & J2032+4127 & 0.28 & 0.77 & 7\\
J2108+5157   &     &     &    &     \\
J2226+6057 & J2229+6114 & 5.89 & 9.38 & 64\\
\enddata

%% Include any \tablenotetext{key}{text}, \tablerefs{ref list},
%% or \tablecomments{text} between the \enddata and 
%% \end{deluxetable} commands

%% No \tablecomments indicated

%% No \tablerefs indicated
\label{tab:ratio2}
\end{deluxetable}

%{\bf Maximum energy}

PWNe have been recognized as one of the most efficient electron factories in our Galaxy \citep{2018A&A...612A...2H}. They are powered by energetic pulsars, which inject ultra-relativistic electrons and positrons in their magnetosphere. These particles form a cold ultra-relativistic wind, expanding with bulk Lorentz factor $\Gamma$ in the range $10^4$-$10^7$, until reaching the termination shock (TS) \citep{1974MNRAS.167....1R,1984ApJ...283..710K}. At the shock, particles are believed to be accelerated to multi-TeV energies, inflating a non-thermal nebula which constitutes the {\it plerion} \citep{amato20}. A large fraction of the pulsar spin-down power, $\dot{E}$, is radiated in the very high energy regime via inverse Compton (IC) scattering, resulting in a power-law spectrum that can extend up to at least a few tens of TeV \citep{amatoolmi21,1997MNRAS.291..162A,2009ASSL..357..451D,2006ARA&A..44...17G, 2022A&A...660A...8B}. At these energies, the scattering occurs mostly in deep Klein-Nishina regime, where electrons lose most of their energy in a single scattering event and the maximum energy observed in photons roughly coincides with the maximum energy to which the electrons are accelerated. Radiation losses beyond hundreds of TeVs, into the PeV regime, are very rapid and demand an extremely efficient acceleration rate.

Indeed the connection between pulsars and PeVatrons provides important constraints, which stand on first principles, independently of more sophisticated modeling. First, the absolute maximum energy the particles can reach, E$_{\rm max}$, depends ultimately on the maximum potential drop between the pulsar and infinity, $\Phi_{\rm PSR}=(\dot E/c)^{1/2}$, with $c$ the speed of light. 
Since the particle acceleration, regardless of the acceleration mechanism, is always carried out by the electric field $|\vec E|$, this maximum energy is related to the electric potential associated with this field. Therefore, E$_{\rm max}$ can be defined in terms of the maximum potential drop and has as an absolute maximum at the value $E_{\rm max}=q (\dot E/c)^{1/2}$, where $q$ is the charge of the particle. This is equivalent to saying that the maximum energy of the particle depends on the size of the accelerator, which in the case of PWNe is the size of the TS, as deduced from X-ray observations of young, well studied systems, thus $E_{\rm max}=q|\vec E|$R$_{\rm TS}$. If the wind is described as an ideal magnetohydrodynamic flow \citep{1984ApJ...283..710K,2014MNRAS.438..278P,2019MNRAS.488.5690O}, the electric field strength cannot exceed the magnetic one B$_{\rm TS}$. Defining the ratio between the two as $\eta_{e}$ (which is $\eta_{e}\le$1 in such ideal conditions), the former equation can be written as $E_{\rm max}=q\eta_{e}$B$_{\rm TS}$R$_{\rm TS}$, which is the well-known {\it Hillas} criterion \citep{1984ARA&A..22..425H}.

%This is equivalent to saying that the acceleration needs to fulfill the {\it Hillas} criterion \citep{1984ARA&A..22..425H}, i.e., the particle Larmor radius should not exceed the size of the acceleration region. \eow{This region, for a magnetohydrodynamic flow, is limited by the radius of the TS, thus} $E_{\rm max}=\eta_{e}$qR$_{\rm TS}$B$_{\rm TS}$, with $\eta_{e}\le$1 (in such flow conditions) defined as the ratio between the accelerating electric field and the magnetic field \wipeouteow{E$_{\rm max}$=qR$_{\rm TS}$B$_{\rm TS}$} %(or $E_{\rm max}=\beta \eta_{e}$qR$_{\rm TS}$B$_{\rm TS}$, with $\eta_{e}= 1$ defined as the ratio between accelerating electric field and magnetic field, and $\beta$=1 the velocity in units of speed of light) where B$_{\rm TS}$ is the local magnetic field strength.
%and B$_{\rm TS}$ the local magnetic field strength. \wipeouteow{ (or $E_{\rm max}=\eta_{e}$qR$_{\rm TS}$B$_{\rm TS}$, with $\eta_{e}$ defined as the ratio between the accelerating electric field and the magnetic field, forced to be $\eta_e\leq1$ in a magnetohydrodynamic flow).}
The magnetic energy density (defined as B$_{\rm TS}^2$/8$\pi$) can be expressed as a fraction $\eta_{\rm B}$ of the pulsar wind energy flux:
\begin{equation}
 \frac{B_{\rm TS}^2}{8\pi} = \eta_{\rm B}\frac{\dot E}{(4 \pi R_{\rm TS}^2 c)}
\end{equation}
%The latter is defined as $\dot E/(4 \pi R_{\rm TS}^2 c)$, 
thus:
\begin{equation}
\begin{split}
B_{\rm TS} & =(2\eta_{\rm B})^{1/2}R_{\rm TS}^{-1} (\dot E/ c)^{1/2}  \\
& = 25 \eta_{\rm B}^{1/2}R_{\rm 0.1}^{-1}\dot{E}_{36}^{1/2}~~\rm{\mu G}\ ,
\label{eq:bts}
\end{split}
\end{equation}
where $R_{0.1}$ is the termination shock radius in units of 0.1pc and $\dot E_{36}$ in units of $10^{36}$ erg/s. Using this expression in the expression of E$_{\rm max}$ described above results in the following limit to the maximum energy of the accelerated particles, independently of whether electrons, positrons or protons: %\sout{(in case of electrons and protons)}}:
\begin{equation}
E_{\rm max}\approx 2\ \eta_e\ \eta_{\rm B}^{1/2}\ \dot E_{36}^{1/2}~~\rm{PeV}
\label{eq:emax}
\end{equation}
The fraction of pulsar wind energy flux transferred to magnetic field is constrained, by energy conservation, to be $\eta_{\rm B}\leq1$. The expression above, which is independent of the particle species, can be used to derive an absolute maximum to the energy to which particles can be accelerated.

For electrons in the multi-TeV regime, the most relevant target for IC scattering is the 2.7~K CMB photons. The photon and electron energy can then be related using the approximation presented by \citealt{2021Sci...373..425L}: $E_{\rm{e}}\simeq2.15 E_{\gamma,15}^{0.77}$~PeV (with $E_{\gamma,15}$ in units of $10^{15}$eV, or PeV), which provides an accuracy better than 10\% above 30~TeV. Thus Eq.~\ref{eq:emax} can be written as: 
\begin{equation}
E_{\rm \gamma~\rm max}\approx 0.9\ \eta_e^{1.3}\ \eta_B^{0.65}\ \dot E_{36}^{0.65}~~\rm{PeV}
\label{eq:gmax}
\end{equation}

The above expression provides a direct link between the observed maximum energy in the gamma-ray spectrum and the spin-down power of the pulsar. It can be immediately derived that only very energetic pulsars with at least $\dot{E}\gtrapprox10^{36}$erg/s could power the observed PeV gamma rays. %
%When considering a more realistic ratio of the
%magnetic to the accelerating electric field of $\eta_B\sim10\%$ \citep{2011ASSP...21..435W}, a larger energy budget of $10^{37}$erg/s has to be invoked to reach such extreme energies.\elena{Here $\eta_B$ is the ratio between magnetic pressure and total pressure, not between electric and magneic field. That would give an extra factor that we are not including here.}  

Additionally, a second condition is required to shine in gamma rays up to PeVs: the acceleration rate $\tau_{\rm acc}$ should also overcome the radiative losses of the parent electrons. The first can be expressed as a function of the magnetic field at the TS as $\tau_{\rm acc}=E_{\rm e}/(\eta_{e} e B_{\rm TS}c)$. The overall cooling time, that can be written as $\tau_{\rm{loss}}=(1/\tau_{\rm sync} + 1/\tau_{\rm IC})^{-1}$, is dominated, even for a few $\mu$G magnetic field, by synchrotron losses, above a few hundreds of TeV. In the Klein-Nishina limit, the IC cooling time of electrons in 2.7 K CMB depends only on the electron energy as $\tau_{\rm IC}\simeq~10^{12}E_{e,15}^{0.7}$~s \citep{2014ApJ...783..100K}. The former can be compared with the synchrotron time $\tau_{\rm sync} \simeq~4\times 10^9E_{e,15}^{-1}B_{-5}^{-2}$~s , where $B_{-5}$ is the magnetic field in units of 10$\mu$G. The condition $\tau_{\rm acc} = \tau_{\rm syn}$ results in the following expression for the maximum energy of the electron population:
\begin{equation}
E_{\rm e~\rm max}\approx 20\ {\eta_e}^{1/2} B_{-5}^{-1/2}~~\rm{PeV}\ .
%E_{\rm e~\rm max}\approx 20{\eta_e}^{1/2} B_{-5}^{-1/2}~~\rm{PeV}
\label{eq:emaxb}
\end{equation}
This corresponds to a maximum energy of the photons produced by IC scattering on the CMB of
%or using the corresponding gamma-ray energy in the IC scenario on the 2.7 K CMB background:
\begin{equation}
E_{\gamma~\rm max}\approx 5{\eta_e}^{0.65} B_{-5}^{-0.65}~~\rm{PeV}\ ,
\label{eq:gmaxb}
\end{equation}
or, in terms of pulsar $\dot E$ and TS radius, using Eq.~\ref{eq:bts}:
%The maximum energy \elena{determined by \sout{related to the}} losses can also be expressed as a function of the spin-down \elena{power \sout{energy}} and \elena{TS radius \sout{termination shock}}, as:
\begin{equation}
 E_{\gamma~\rm max}\approx 2.7
 {\eta_e}^{0.65}{\eta_{\rm B}}^{-0.33}R_{\rm 0.1}^{0.65}\dot{E}_{36}^{-0.33}~~\rm{PeV}\ .  
 \label{eq:gmaxrts}
\end{equation}

%This equation can be used to put \elena{an upper \sout{a lower}} limit on the magnetic field strength at the TS:
%\begin{equation}
%\begin{split}
%    B_{\rm{TS}} = 4{\eta_e}E_{\rm e~\rm max}^{-2}~~\rm{mG} \\ 
%= 0.8{\eta_e}E_{\gamma~\rm %max}^{-1.54}~~\rm{mG}. 
%\label{eq:btsmaxg}
%\end{split}
%\end{equation}
The comparison between Eq.~\ref{eq:emax} and \ref{eq:emaxb} shows that radiation losses pose the most serious challenge to reach PeV energies only for young, energetic pulsars ($\dot{E}	\gtrsim 10^{37}$erg/s), with a magnetic field at the wind TS in the 100$\mu$G range, such as the Crab Nebula \footnote{Note that in scenarios where acceleration and radiation do not occur in the same region, the absolute maximum energy is still determined by the potential drop.}.
%\eow{I am moving the sentence starting with "Out of all pulsars..." to the next section - here we have not yet discussed any association with the pulsars - I also tried to summarise your paragraph on the crab (below!)}
This does not mean that the Crab is a poor accelerator. On the contrary, as discussed in \citealt{2021Sci...373..425L}, the Crab Nebula, with an estimated magnetic field of $\approx112~\mu$G, requires an acceleration rate corresponding to $\eta_e\approx0.16$ to reach the energies observed (E$_{\gamma,\rm{max}}=1.1$~PeV). These large values of $\eta_e$ are impressive, being several orders of magnitude larger than those inferred for other powerful accelerators, such as e.g. supernova remnants (SNRs), for which typically $\eta_e\approx 10^{-3}$ \citep{2001RPPh...64..429M}. Even though larger values of $\eta_e\geq1$, are possible during e.g. gamma-ray flares \citep{blandbuhler14}, the large value derived in Crab only reflect the relativistic nature of the acceleration, which in such shocks is far from being understood (see e.g. \citealt{amato20, amatoolmi21}).

The constraints so far discussed are related to the maximum photon energy observed. Additional information can be obtained by comparing the IC luminosity of the PeV sources and the total power injected by the pulsar in its surrounding. The total energy in electrons responsible for IC radiation can be derived using the gamma-ray observations as $W_{\rm e,\gamma} = L_{\gamma}\tau_{\rm IC}$. $W_{\rm e,\gamma}$ cannot exceed $W_{\rm{e,PSR}}$, the total energy made available by the pulsar in the form of gamma-ray emitting electrons. Since the lifetimes of the latter are determined by losses, one can write $W_{\rm{e,PSR}} = \gamma_{\rm eff} \dot{\rm E}\tau_{\rm loss}$, with $\gamma_{\rm eff}$ the fraction of $\dot E$ converted into gamma-ray emitting electrons:

%\sout{whereas the total energy in electrons powered by the rotational energy $\dot{\rm E}$ can be expressed as $W_{\rm{e,PSR}} = \dot{\rm E}\tau_{\rm loss}$. The apparent conversion efficiency, $\gamma_{\rm eff}$, can be thus expressed as:}

%\begin{equation}
%\begin{split}
%\gamma_{\rm eff} & = \frac{L_{\gamma}}{\dot{E}} \left(1+ \frac{\tau_{\rm IC}}{\tau_{\rm syn}}\right) \\
%& = \frac{L_{\gamma,36}}{\dot{E_{36}}} (1 + \eow{125 \ \sout{260}}E_{e,15}^{1.7}B_{-5}^{2})
%\label{eq:efficiency}
%\end{split}
%\end{equation}
\begin{equation}
\begin{split}
\gamma_{\rm eff} & = \frac{L_{\gamma}}{\dot{E}} \left(1+ \frac{\tau_{\rm IC}}{\tau_{\rm syn}}\right) \\
& = 10^{-4}\frac{L_{\gamma,32}}{\dot{E}_{36}} (1 + 260\ E_{e,15}^{1.7}B_{-5}^{2})
\label{eq:efficiency}
\end{split}
\end{equation}
which also depends strongly on the magnetic field. The maximum allowed conversion efficiency from rotational power to gamma-ray one ($\gamma_{\rm eff}$ = 1) results in another boundary to pulsars as ultra-high energy sources. %{\color{orange}However, when evaluating the capability of a energetic pulsar to power an ultra-high energy sources, this condition is in general less constraining than the ones related to the maximum energy, given the large energy reservoir injected by pulsars.} 

% I just wanted to indicate that in many occasions we dont find lots of constrains related to the IC luminosity when looking to energetic pulsars - but I guess that's implicit when selecting 1e36 pulsars - I removed it 

%The maximum allowed conversion efficiency from rotational power to gamma-ray one ($\gamma_{\rm eff}$ = 1) results in another constraint to the maximum magnetic field. For a given energy in electrons E$_{e,15}$:  
%\begin{equation}
%B_{\rm TS}\approx8\times10^{-4}E_{e,15}^{%-0.85}(\dot E_{36}/L_{\gamma, %36})^{1/2}~~\rm{mG}
%\label{eq:bmax}
%\end{equation}
 
% Following the argument above and substituting here equation \ref{eq:bts}, we can also relate the maximum energy to the TS using the gamma-ray luminosity:
 
% \begin{equation}
%     E_{\rm{e~max}} = 0.02\eta_{B}^{-0.6}L_{\gamma,36}^{-0.6}R_{0.1}^{1.2}\gamma_{\rm eff}^{0.6}  ~~{\rm PeV}
% \end{equation}
 
 %\rlc{I think that the 0.02 is missing the exponent (1/0.85=1.2)}

\begin{figure*}%[t]
  \centering
  \includegraphics[width=\textwidth]{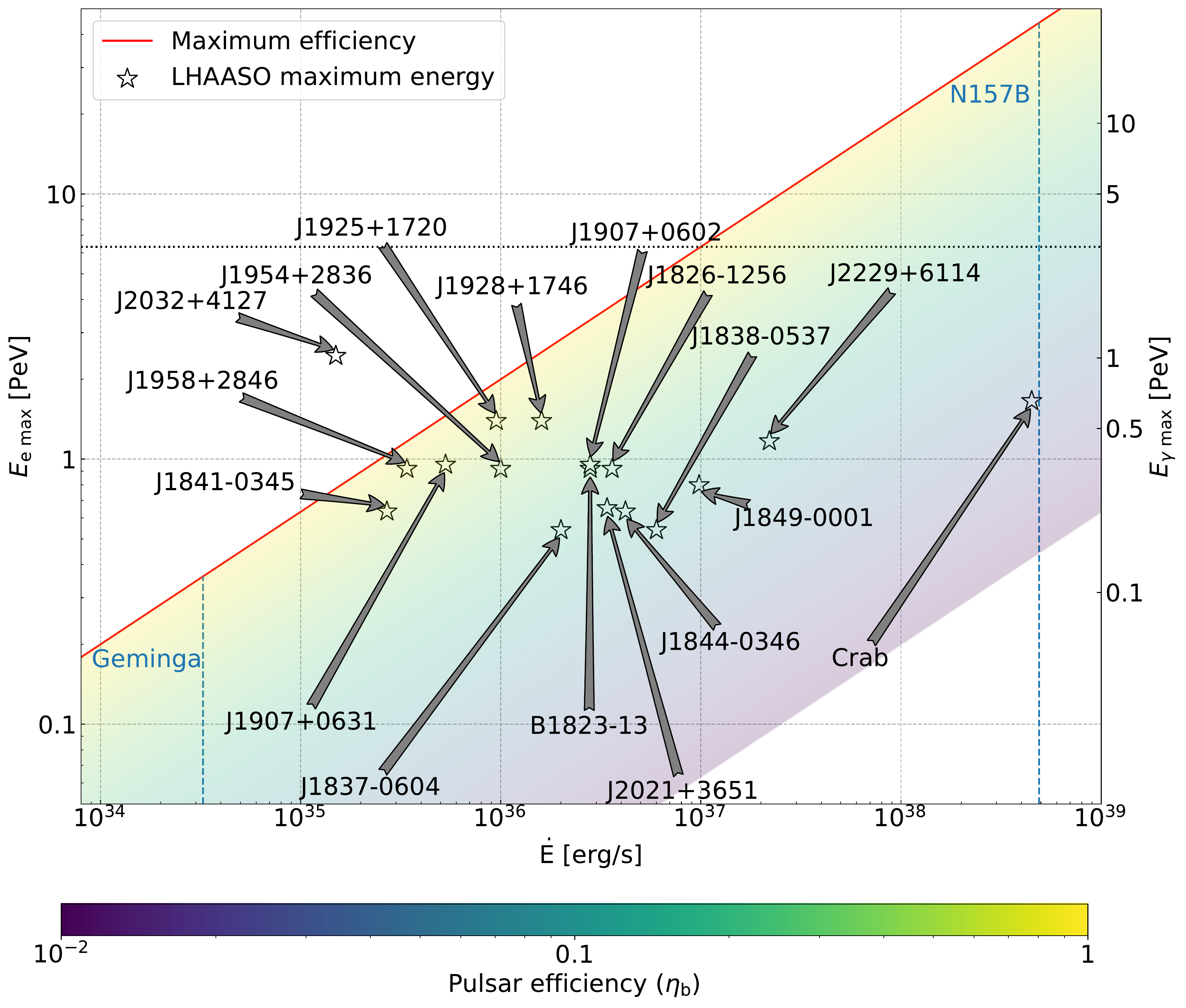}
  \caption{Maximum electron energy derived from the LHAASO spectra versus spin-down power of the co-located pulsars. The right Y-axis shows the corresponding gamma-ray energy. The colored area shows the values for $\eta_e\ \eta_B^{1/2}$ ranging from 0.01 to 1, with the red line indicating the limiting value corresponding to maximally efficient acceleration $\eta_e=1$ and $\eta_{\rm B}=1$. The dotted black line marks the upper limit to the maximum energy for young pulsars with large magnetic field of 100~$\mu$G. The blue dashed horizontal lines show the predicted values for PWNe associated to Geminga and N157B}.

  \label{fig:Ee_vs_edot}
\end{figure*}

\begin{comment}

\begin{figure}%[t]
  \centering
  \includegraphics[width=0.5\textwidth]{figures/Efficiency_vs_edot.pdf}
  \caption{Efficiency to transform spin-down power into VHE $\gamma$-ray emission above 100 TeV.}
  \label{fig:Efi_vs_edot}
\end{figure}
\end{comment}

%%%%%%%%%%%%%%%

\section{Comparison with pulsars in the region of interest}

Twelve ultrahigh-energy gamma-ray sources were reported by \citealt{2021Natur.594...33C}, with a spectral energy distribution extending up to more than 100 TeV, one of them being associated with the Crab Nebula. The maximum energy in photons E$_{\rm \gamma\ LHAASO}$ and electrons E$_{\rm e\ LHAASO}$, derived from the LHAASO observations, are listed in Table 1. The latest was derived from the photon energy using the expression in \citealt{2014ApJ...783..100K}. The majority of these sources show a diffuse gamma-ray structure, with angular extensions up to 1$^{\rm o}$. This extended structure makes the association with the PeVatron accelerator complex. To explore the possibility of an association of the ultra-high-energy sources with pulsars, we searched for relatively young ($\tau < 10^6$ yrs), energetic ($\dot{E}$/d$_{\rm kpc}^2 >10^{34}$ erg/s/kpc$^2$, or $\dot{E}>10^{36}$ erg/s when the distance is unknown) pulsars in the ATNF catalog\footnote{http://www.atnf.csiro.au/research/pulsar/psrcat/}, located within 1$^{\rm o}$ around the position of the LHAASO sources. %The selection criterion is motivated by the general trend observed on the TeV PWNe population, for pulsars powering bright multi-TeV sources \citep{2018A&A...612A...2H}. %The ratio $\dot{E}$/d$_{\rm kpc}^2$ also provides an indication of the level of the expected gamma-ray flux for a given pulsar with spin-down energy $\dot{E}$, located at a distance d$_{\rm kpc}$ (see \citealt{1997MNRAS.291..162A}). 

For each of the LHAASO sources, we found at least one pulsar (two in some cases) which could potentially be linked to it, except for the source J2108+5157, for which no bright pulsar is found in the vicinity. The selected pulsars and their properties are listed in Table 1. The LHAASO source associated to the Crab Nebula has been described in detail in \citealt{2021Sci...373..425L}.

To evaluate the potential of an associated pulsar to power the LHAASO sources, we estimate the maximum energy to which particles can be accelerated by such pulsar, assuming $\eta_B=\eta_e=1$ in Eq.~\ref{eq:emax}. The results are summarised in Fig.~\ref{fig:Ee_vs_edot} and the maximum energies of the particles and corresponding gamma-ray E$_{\gamma~\rm{max}}$, in the case of IC scattering of the CMB, are listed in Table 2. In Fig.~\ref{fig:Ee_vs_edot} we place the potential associations in the $\dot E$-$E_{\rm e,max}$ plane and compare them with theoretical predictions based on Eq.~\ref{eq:emax} for $\eta_e\ \eta_B^{1/2}$ ranging from 0.01 to 1. %\sout{We compare the energy in particles achievable by a given spin-down power $\dot{E}$, assuming a magnetic conversion efficiency $\eta_{\rm B}$ ranging from 0.01 to 1.}
Out of all pulsars possibly associated to the 12 LHAASO sources, only in the Crab pulsar the maximum energy is limited by the radiation losses%\eow{(if the particle acceleration and loses occur in the same region)}
, while for all other pulsars the most relevant constraint will come from saturation of the full available potential drop (Eq.~\ref{eq:emax}). %\eow{Note that synchrotron losses only dominates for young, highly magnetized ($\sim100\mu G$), with large spin-down luminosity pulsars $>10^{37}$erg/s.}
The upper limit to the maximum electron (and photon) energy, using $\eta_e=1$ and 100$\mu G$, is marked with a dotted horizontal line in Fig. \ref{fig:Ee_vs_edot}. %\wipeouteow{The red line indicates the largest possible value 
%of $\eta_{\rm B}=1$, 
%of the electron} %\wipeouteow{and photon energies, corresponding to maximally efficient} \wipeouteow{acceleration and the entire wind energy flux ending up} \wipeouteow{ in magnetic field}
Above the red line, the particle flow would require values of $\eta_B=1$ and $\eta_e>1$, and would demand non-ideal mechanisms (see \citealt{amatoolmi21} for a review). %Each of the pairs ultra-high-energy source/energetic pulsar is indicated by a star. 
Only one of pairs ultra-high-energy source/energetic pulsar (LHAASO~J2032+4102 / J2032+4127) lies above the absolute maximum, resulting in an impossible connection between the two (if the spin-down power of the pulsar is correct within a factor of $\sim$4). We also marked in Fig.~\ref{fig:Ee_vs_edot} two remarkable pulsars with vertical blue lines: Geminga and the Crab twin, N157B, located in the Magellanic Cloud. With a moderated spin-down luminosity of $3.26\times10^{34}$erg/s, but located at a small distance of 250 pc, the Geminga Nebula is a prime target for LHAASO, given its large size of $\sim$2$^{\rm o}$ \citep{2017Sci...358..911A}. N157B is, on the contrary, the furthest gamma-ray PWN detected \citep{2012A&A...545L...2H}, but its large spin-down luminosity ($\dot{\rm E}=4.9\times10^{38}$erg/s) and similarity with the Crab Nebula makes it also an interesting source to understand the contribution of pulsars to the PeV sky.

We can also use the spectral parameters of the PeV sources, in particular E$_{\gamma~\rm{max}}$ and gamma-ray luminosity to impose an upper limit on the magnetic field. Constraints are provided by the fact that synchrotron losses should not forbid acceleration up to E$_{\rm e~max}$ (Eq.~\ref{eq:emaxb}) and the energy input from the pulsar be sufficient to power the gamma-ray source (Eq.~\ref{eq:efficiency}).%The magnetic field should be such that the acceleration rate dominates the synchrotron losses, and also the luminosity in gamma rays is sufficient to explain the observed source.
%The limits obtained using Eq. \ref{eq:btsmaxg}, even using $\eta_{\rm e}\approx0.16$ (as the one derived for the Crab Nebula) are not very constraining ($\sim$a few tens of mG), in comparison with the ones obtained using Eq. \ref{eq:bmax}. The latest values are also included in Table \ref{tab:ratio2}. 
We found that, in general, the latter constraint is much stronger and requires that the magnetic field cannot exceed a few tens of $\mu$Gauss, which agrees with the typical values derived from very-high energy observations in the TeV regime (see e.g. \citealt{2018A&A...612A...2H}).  Despite these low constraints in the magnetic field, the Larmour radius of the electrons with the highest energies is still in agreement with the typical size of the TS, defined by the balance between the wind pressure and the one from the surrounding medium (see e.g.\ \citealt{2008AIPC..983..171K}).
%{\color{orange} I commented this it out because I moved this info above.}
%Note that for young, very energetic pulsars with spin-down luminosity larger than $10^{37}$erg/s like the Crab pulsar or N157B, and with magnetic fields of the order of $\sim100\mu G$, the maximum energy is limited by radiation losses. 

%In addition, the  efficiency $\gamma_{\rm eff}$, obtained using Eq. \ref{eq:efficiency} with a magnetic field value of the order of a few $\mu$G, is for all these pulsars $\gamma_{\rm Eff}\le 10^{-2}$. This is also in line with those derived using TeV observations of PWNe \citep{2018A&A...612A...2H}.}
%{\bf do we want to add anything about the TS? something like: TS $\sim$ 0.05 - 0.1 provides $B_{ts}$ compatible with the B upper limit? }

\section{Concluding Remarks}

 We derived the absolute maximum energy that can be accelerated by pulsars, obtained from the maximum potential drop available, without further assumptions beyond ideal MHD flow. This maximum energy can now be confronted with observational results as those recently published by LHAASO. The extreme energies reached in the sources detected by LHAASO provide direct information about current particle acceleration, given the fast cooling time involved beyond hundreds of TeV. At these energies, the up-scattering of the 2.7 K CMB radiation dominates the observed gamma-ray radiation in PWNe, providing a powerful diagnostic tool. Additionally, these multi-TeV electrons propagating in the magnetised nebulae should also power an X-ray nebula, visible at a few keV. %:
%\begin{equation}
%\epsilon \simeq 10 (E_\gamma/100~\rm{TeV})^{1.5}(B/3~\mu %\rm{G})~~\rm{keV} \label{eq:esyn}   
%\end{equation}
The detection of such an extended nebula is challenging for pointing X-ray instruments like XMM-{\it Newton} or {\it Chandra} \citep{2008AIPC..983..171K,2019ApJ...875..149L}. However, the new X-ray satellite eROSITA, sensitive to X-rays in the energy range of 0.3 - 11 keV and with a wide field of view of 0.81$^{\rm o}$, is optimal to constrain the X-ray counterpart. The expected sensitivities achieved by
eROSITA for extended sources in the energy range of 0.5-2 keV are
$1.1\times10^{-13}~\rm{erg}/\rm{cm}^{2}/\rm{s}$ for the first all-sky survey (eRASS:1), and $3.4\times10^{-14}~\rm{erg}/\rm{cm}^{2}/\rm{s}$ for the four year all-sky survey (eRASS:8). With fluxes at 100~TeV ranging from $\sim 5 \times10^{-13}$ erg/cm$^2$/s to $\sim 5\times10^{-12}$ erg/cm$^2$/s, the  X-ray counterpart, assuming a magnetic field as low as 3$\mu$G, should be expected with fluxes above $\sim 2 \times10^{-12}$ erg/cm$^2$/s in the 0.5 to 2 keV band, an order of magnitude larger than the eROSITA sensitivity. These numbers should be taken with caution, since the surface brightness might not be homogeneous across the large TeV source region. 

From the 11 sources considered, two sources stand out, and different accelerators and/or gamma-ray production mechanisms should be investigated: LHAASO\,J2108+5157 and\,LHAASO J2032+4102. The first is extensively discussed in \cite{2021arXiv210609865T} %It is located above the Galactic plane at a latitude of $\sim$3$^{\rm o}$, 
and it is found to be point-like, within the angular resolution of LHAASO for this analysis (0.26$^{\rm o}$). 
The closest pulsar is $\sim3^{\rm o}$ away, which at 2-3 kpc corresponds to more than 100\,pc away. No counterpart has been found neither in the TeV regime. %In the GeV regime, an extended source lies within the PeV region, but the association between this and the LHAASO source is not clear yet \citep{2021arXiv210609865T}.  
The second source is co-located with a pulsar (PSR\,J2032+4127) in an interacting binary system \citep{2018ApJ...867L..19A,2019arXiv190804165W}. The above considerations only apply to the isolated pulsar, however the mixing of the two winds could in principle  lead to different conclusions. %Despite being young and energetic, electrons can only be accelerated up to a maximum energy of 0.7\,PeV, which is not sufficient to explain the extension of the LHAASO spectrum, corresponding to energy in electrons up to 2.45\,PeV. 
%This pulsar has been previously associated to the first-discovered TeV unidentified source (TeV\,J2032+4130, \citealt{2002A&A...393L..37A,2007ApJ...658.1062K,2008ApJ...675L..25A,2014ApJ...783...16A}) reported by the HEGRA collaboration in 2002. 
The pulsar powers a compact $\sim$0.2$^{\rm o}$ gamma-ray nebula \citep{2002A&A...393L..37A,2007ApJ...658.1062K,2008ApJ...675L..25A,2014ApJ...783...16A}, which can only partially be connected to the very extended ultra-high-energy source. %{\bf Recently, radio and TeV observations have revealed that the pulsar is in orbit with a Be massive star \citep{2018ApJ...867L..19A,2019arXiv190804165W}.}.
The system is located at the heart of the Cygnus cocoon, a bright GeV and TeV extended diffuse emission, which has also been connected with several individual sources, including the massive stellar cluster Cygnus OB \citep{2011Sci...334.1103A,2021NatAs...5..465A,2019NatAs...3..561A}. The potential connection between the Cygnus cocoon and the LHAASO source opens interesting prospects for stellar clusters as contributors of ultra-high energy particles \citep{bykovsfr}. 

Further information regarding the morphology of these sources should provide crucial insight into the origin of the emission. Indeed, electrons and positrons at these energies undergo fast losses due to synchrotron radiation, and might appear as compact, sub-degree regions. However, for low enough (a few $\sim\mu$G) magnetic fields, and fast enough transport, these electrons could still fill up a volume larger than a few tens of parsecs, which would match the large extension observed, if located close enough to us (few kpc). Alternatively, these electrons might have escaped into the interstellar
medium, filling up a {\it halo} where particles are essentially free from their parent PWN \citep{2022NatAs...6..199L}. This is particularly relevant for nebulae like HESS\,J1825--137, for which a clear energy-dependent morphology has been established in the TeV regime \citep{2019A&A...621A.116H}. Observations above 100~TeV should provide a clear picture of the radiative cooling and propagation of electrons. Note that regardless the propagation regime, the maximum energy to which particle can be accelerated is always limited by the equations derived here. Gamma-ray images at different energies should also serve as test-bench for other effects involving ballistic and diffusive propagation, which could play an important role in the observed morphology \citep{2015PhRvD..92h3003P}.

\bibliography{ms_v1}{}
\bibliographystyle{aasjournal}



\end{document}